# Virtual Mouse And Assistant: A Technological Revolution Of Artificial Intelligence

Jagbeer Singh , Yash Goel, Shubhi Jain, Shiva Yadav

Dept. of Computer Science and Engineering
Meerut Institute of Engineering and Technology Meerut, India
jagbeer.singh@miet.ac.in, yash.vikas.cs.2019@miet.ac.in, shubhi.jain.cs.2019@miet.ac.in, shiva.yadav.cs.2019@miet.ac.in,
**DOI:** 10.47750/pnr.2022.13.S10.362

## Abstract

The purpose of this paper is to enhance the performance of the virtual assistant. So, what exactly is a virtual assistant. Application software, often called virtual assistants, also known as AI assistants or digital assistants, is software that understands natural language voice commands and can perform tasks on your behalf. What does a virtual assistant do. Virtual assistants can complete practically any specific smartphone or PC activity that you can complete on your own, and the list is continually expanding. Virtual assistants typically do an impressive variety of tasks, including scheduling meetings, delivering messages, and monitoring the weather. Previous virtual assistants, like Google Assistant and Cortana, had limits in that they could only perform searches and were not entirely automated. For instance, these engines do not have the ability to forward and rewind the song in order to maintain the control function of the song; they can only have the module to search for songs and play them. Currently, we are working on a project where we are automating Google, YouTube, and many other new things to improve the functionality of this project. Now, in order to simplify the process, we've added a virtual mouse that can only be used for cursor control and clicking. It receives input from the camera, and our index finger acts as the mouse tip, our middle finger as the right click, and so forth.

**Keywords**—Python, OpenCV, Hand Gesture, speech recognition.

## I. INTRODUCTION

Technology advancements have had a wide range of goods on day-to-day living. The way we lived in the past and how we live now are veritably different as a result of technological advancements. Prior to the development of machine learning, artificial intelligence, and other technologies, computers were only able to perform a small number of tasks. As a result, a number of voice assistants, such as Alexa, Siri, and Cortana, have been developed. These voice assistants use speech recognition to improve computers and do away with the need for input devices.

There are many more devices that are part of the AI virtual assistant. like Amazon, Google, and Microsoft. In the new model of Amazon's Echo virtual assistant, the user wakes up or uses the device by saying "Alexa." Then a light appears on it, which indicates that it is ready to take commands and proceed. You can give different commands to it, like "play music on Amazon Music or YouTube," "open Netflix," "what is today's temperature," and many more.

Virtual assistant technologies need a lot of data to function since they feed artificial intelligence (AI) platforms, including machine learning, speech-recognition, and natural language processing. Artificial intelligence programming uses sophisticated algorithms to learn from the data provided by the user and improve its ability to predict the user's needs when they interact with a virtual assistant.

The virtual mouse system's main objective is to replace the physical mouse with hand gestures for cursor control. The described system may be implemented via a webcam or built-in camera that identifies hand movements and hand tips and analyses these frames to perform specific mouse activities. The results of the study reveal that the proposed AI virtual mouse system has performed remarkably well, has less delicacy than the current models, and successfully overcomes the bulk of the shortcomings of the latter. The virtual mouse is useful for regular use. Additionally, it will be helpful in instances like COVID outbreaks since it eliminates direct touch.

## II. LITERATURE REVIEW

To lead the concluding project on Virtual Assistant requires a lot of previous research work which enhances my knowledge for completion of the project. Research work concluded as follows:

- ➢ Similar to the comparison of Real-Time AI System for a Virtual Mouse Based on Deep Learning Avoiding COVID-19 Spreading various actions without the usage of a mouse or keyboard demonstrates how science and technology are becoming more powerful every day. Without using a physical mouse, the AI virtual assistant greatly benefits from the use of hand gestures to carry out a variety of operations involving left and right clicks, scrolling, and computer cursor capabilities.

- ➢ Development of CNN based Virtual Mouse has the technique which is used to capture framework by the help of webcam which capture frameworks and function according to the framework easily. This makes it less dependent on the mouse for the input command to the system.

- ➢ Study of Voice Controlled Personal Assistant Device helps and gives the knowledge of the human voice with the technology interaction which helps a lot to the human to complete any work with the ease and comfort. Voice is the power of humans and this power now also matches with the technology to complete the technology task by saying their task with the voice.

- ➢ A real time hand movement recognition system using motion history image gives the guidance to know the sensitivity of hand gesture to use the hand gesture functioning properly and in a restricted manner. This research work teaches the four groups of haar-like directional patterns of hands like up, down, left and right movement of hands.

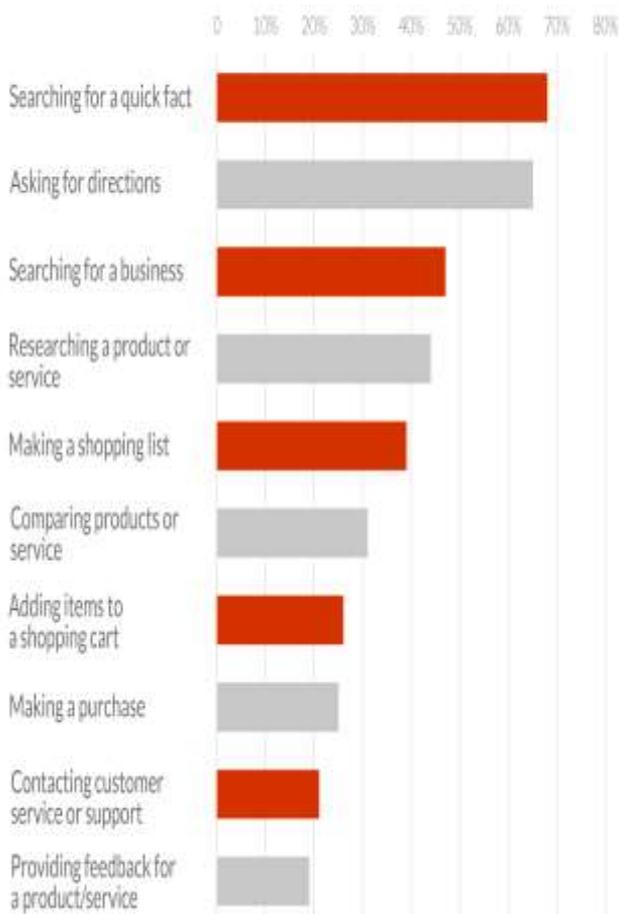

### III. TECHNICAL OVERVIEW

The purpose of this project is to update the virtual assistant by adding Virtual Mouse and add automations on Virtual assistant.

*A.* Voice Assistant

Light (Voice Assistant) works with the help of following modules such as speech recognition, natural language processing and speech synthesis for the result which users want. On the basis of the user command voice assistant listens to the user voice persistently, then gives output to that particular voice in the form of text, after the text it performs the given command and runs accordingly which satisfies the user.

In today's scenario Voice Assistant includes Amazon Alexa Wall Clock (single device application), Google assistant (multiple device application), Cortana (window application), Watson (IBM application), Siri (iPhone application).

Different assistants provide certain specific features in their assistant application which handle situations easily and timely. It resolves many problems of the users.

**Flowchart of voice Assistant**:

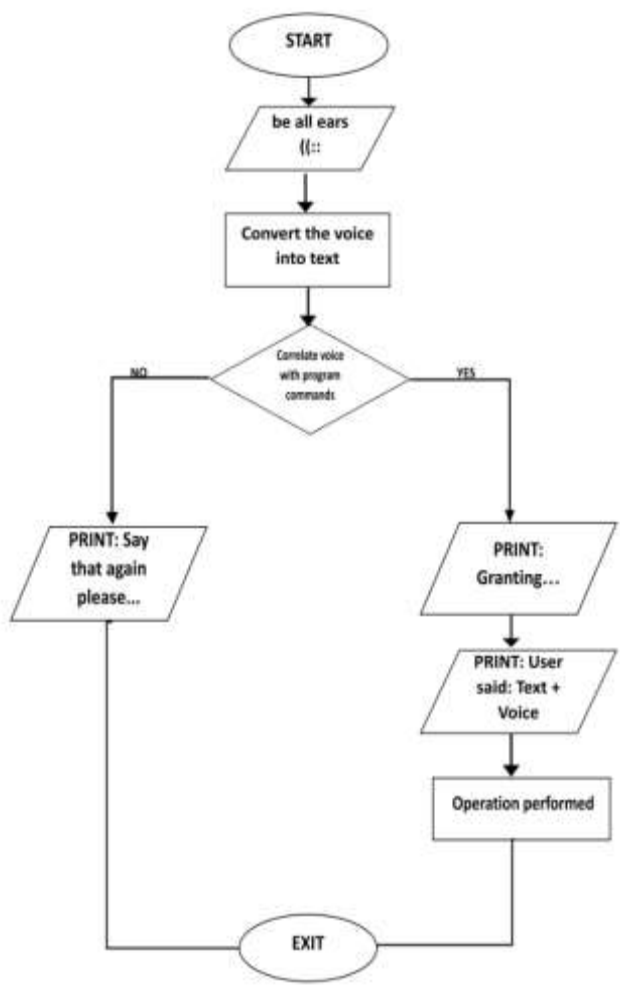

*B.* **Virtual Mouse using Hand Gesture**

Through the use of Web Camera and pyautogui module we do many mouse operations. It is possible to virtually control all mouse operations using hand gestures, along with the assistance of a voice assistant. While doing mouse operation using hand gesture first module is use OpenCV to open a webcam and captures the hands in real-time of system this process is performed to get the real-time view of hand then using hand tracking module place the landmarks point on the hand and done mapping on the hand then using pyautogui we can do any operation virtually through mouse. The hand gestures and assistant works automatically without the help of any external hardware by using "Computer Vision" and "AI ML" algorithms.

## Flowchart of Virtual Mouse:

Various virtual assistant are already present for example Siri that is apple's virtual assistant, Alexa of Amazon and Cortona of Microsoft. For our project we are using the name Friday for our virtual assistant.

Siri:- Siri was developed by Dag Kittlaus and his team at SRI International as his iPhone app. Then in 2010 he introduced Siri by Apple. Siri and its technology are built into the iPhone, iPod Touch, and iPad.

Cortana: Cortana is also a virtual assistant developed by Microsoft that uses the Bing search engine to perform tasks such as answering questions and setting reminders.

Google Now: Google Now was introduced by Google in 2012. Found in a Google search and used in a variety of ways.[15]

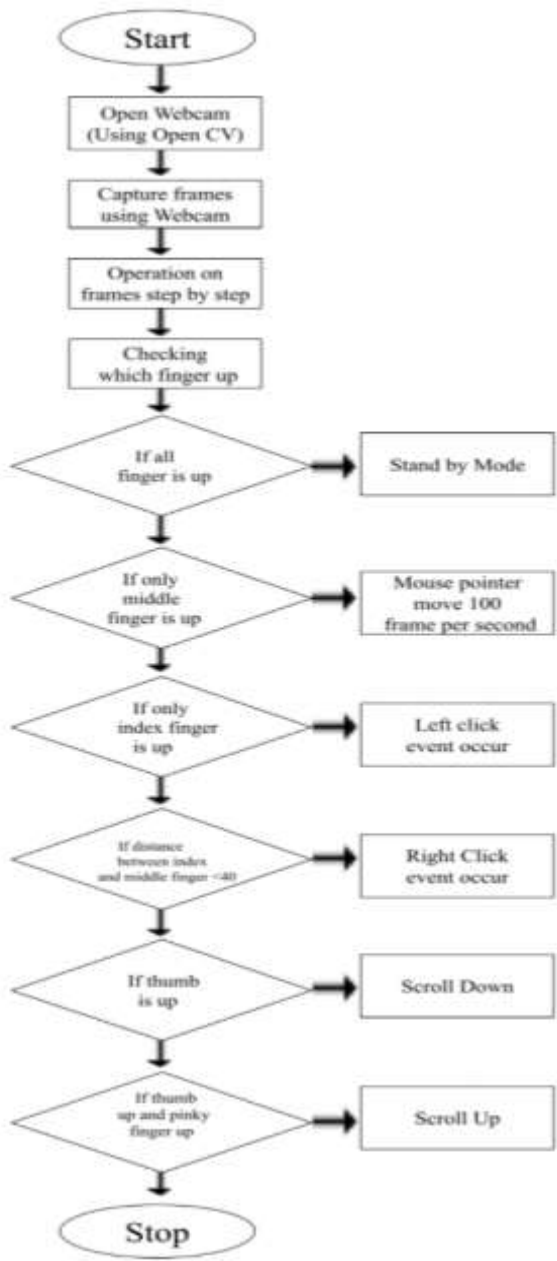

## IV. METHODOLOGY

Need to import required library for making this program like:

- ➢ **CV2:** Its function is to open the Webcam and also to take real-time video.
- ➢ **HandTrackingModule:** Hand tracking is the method by which a computer recognizes a hand from an input image, keeps track of the hand's movement and location, and also marks a landmark point on the hand [15] [16] [17].

- **Pyautogui:** The automation library for Python called PyAutoGui supports keyboard and mouse control. Alternately, we could say that it makes it easier for us to automate the use of the keyboard and mouse to create interactions with other applications when we utilize Python scripts [18] [19] [20] [21] [22].
- **Autopy:** Python's AutoPy is a straightforward, cross-platform GUI automation framework. It has cross-platform, efficient, and simple keyboard and mouse controls, as well as controls for locating colors and bitmaps on the screen and displaying alarms [23] [24] [25] [26] [27].
- **Pyttsx3:** pyttsx3 is a platform-independent, cross-platform text to voice library.
- **Speech_recognition:** Speech_recognition refers to a machine's ability to hear spoken words and identify them.
- **Pywhatkit:** Its function is to search on google and YouTube.
- **Pyscreenshot:** Its function is to take a screenshot.
- **Screen_brightness_control:** Its function is to control brightness of a system.
- Etc.

## VOICE ASSISTANT:

First it receives input(voice) given by the user through speech recognition module then converts the voice input into text by using pytxx3 module then that text matches with the conditions provided in the program if conditions match with program then it executes the operation otherwise does not execute anything [28] [29] [30] [31] [32].

Now we add some automation in Voice assistant for example "YouTube automation in YouTube auto we add features like play music, pause music, fast music, slow music, back music, forward music, full screen", etc.

While performing all these functions a module is needed i.e., Pyautogui (mouse and keyboard controllable automation library). It compresses YouTube shortcut keys.

## VIRTUAL MOUSE:

Virtual mouse is a mouse who completes its task without touching a touch pad. Through this the user needs to put his hand in front of the camera to run the mouse. Firstly, camera capture the real time hand image and after that, that image passes into the hand tracking module then hand tracking module makes the landmarks on hand and do the index(numbering) on fingers then by using fingers up function it check which finger is up according to hand and then finger movement mouse perform its operations [33] [34] [35].

Operations are:

- Mouse Move
- Right Click
- Left Click
- Scroll Up
- Scroll Down

While doing mouse operation using hand gesture first module is use OpenCV to open a webcam and captures the hands in real-time of system this process is performed to get the real-time view of hand then using hand tracking module place the landmarks point on the hand and done mapping on the hand then using pyautogui we can do any operation virtually through mouse. The hand gestures and assistant works automatically without the help of any external hardware by using "Computer Vision" and "AI ML" algorithms [36] [37] [38].

## V. OUTPUT SCREENS

1) No action performed on the screen

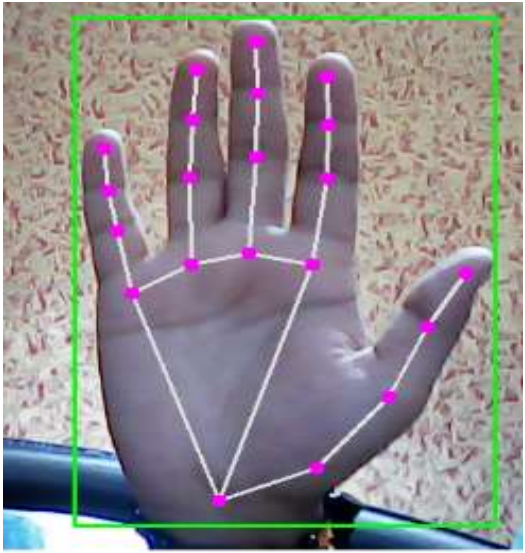
2) For Cursor of mouse to move around the window

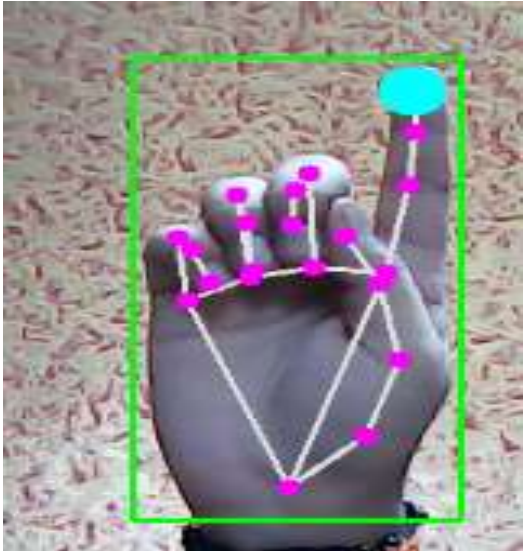
3) Perform right click button with Virtual Mouse Virtual assistants use text as well as speech recognition to make the machine perform different tasks for human. Virtual assistant are coded and are software programs that assist you to perform all your task like a personal assistant for example sending texts, messages, calling contacts.

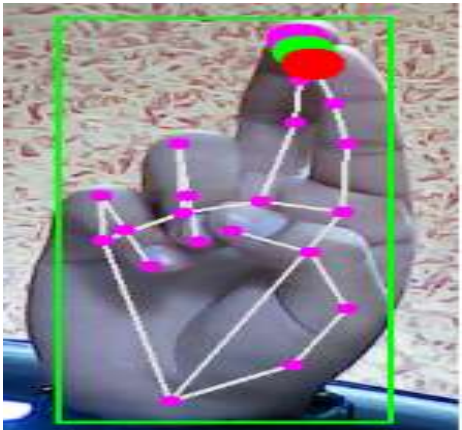
4) Perform Scroll down Operation with Virtual Mouse

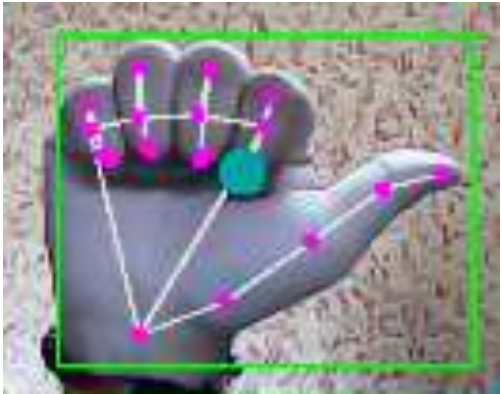

5) Perform Scroll up Operation with Virtual Mouse

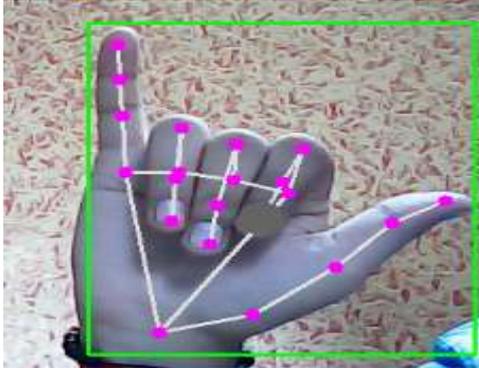

6) For Assistant to Perform Increase Brightness Operation

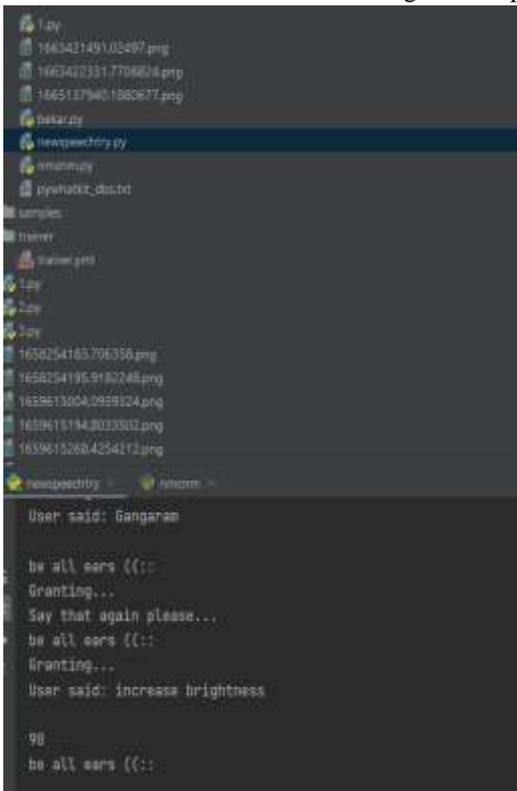

7) For Assistant to Perform Decrease Brightness Operation

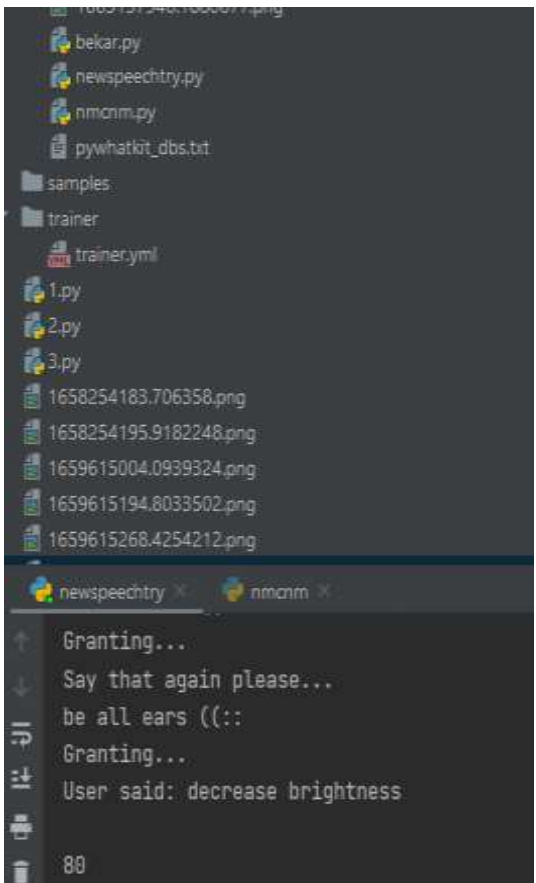

8) For the Assistant to Perform the Temperature Function

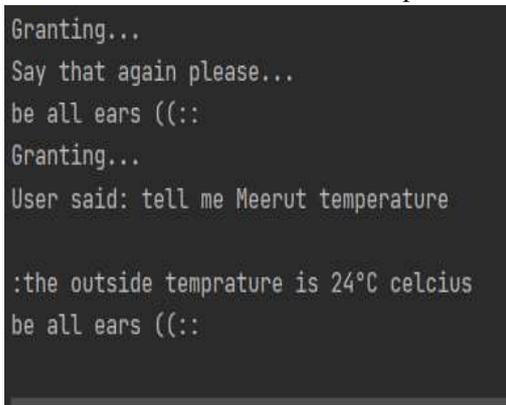

9) For the Assistant to Perform YouTube Automation

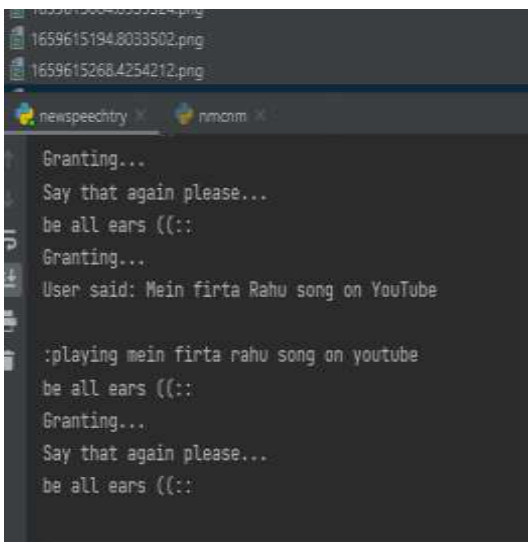

## VI. RESULT ANALYSIS

As we all know the majority of future technological developments will be dominated by Virtual assistant apps because to their usefulness and efficiency. So, we add new features which are not present in the existing virtual assistant. Firstly, we identified who is using Cortana during our research. Over 150 million people are said to be using Cortana, but are they really using Cortana as a voice assistant or just using Cortana Box to do searches in Windows 10. According to the general consensus, Cortana isn't at all useful. You might not notice much of a difference, though, if you primarily used Cortana for work-related tasks like managing your calendar and using Microsoft programmes. Cortana isn't helpful for the typical user. Consequently, we made an effort to create a user-friendly virtual assistant that can perform a variety of functions, such as YouTube Automation, Google Automation, capturing screenshots, telling the battery status, and many more. As a result, we succeed in it. Our algorithm currently has an accuracy of 80-85% for assistance.

The idea of using computer vision to improve human-computer interaction is presented in the proposed AI virtual mouse system.

The limited number of datasets available makes inter-comparison testing of AI virtual mouse systems difficult. Different settings were used to test hand movement tracking and fingertip detection. To summarise the results presented in the table, an experimental test was carried out.

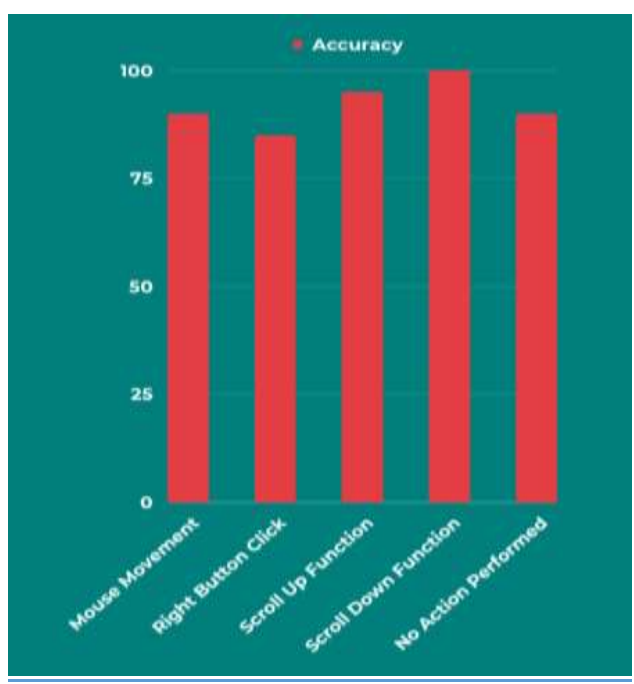

Therefore, our project's accuracy is 75% on average.

## VII. CONCLUSION

In this project we are working with the two tools i.e., voice assistant and virtual mouse. We try to make an efficient voice assistant which completes tasks with at least time. In this we added some automation like Google Automation, YouTube Automation to make fully virtual assistants which users used easily and effectively. With regard to the Virtual mouse, we conclude that we prepared that mouse which works with our fingers without using a touchpad or the mouse ball. By using it we complete all the required work with this like clicking, scrolling, moving cursor, etc. Python 3.7 and 3.9 and open-source modules are used in the development of this project, making it suitable for future updates